\documentclass[aip,rsi,reprint,graphicx]{revtex4-1}
\draft

\usepackage{color}
\usepackage{amsmath}
\usepackage{graphicx}
\usepackage{float}
\usepackage{amsfonts}
\usepackage{amssymb}
\usepackage{amscd}

\begin{document}

\title{Deep underground rotation measurements: GINGERino ring laser gyroscope in Gran Sasso}

% 
% 
% \author{N.~Beverini$^{1,2}$, F.~Bosi$^3$, G.~Carelli$^{1,2}$, D.~Cuccato$^{3,4}$, G.~De~Luca $^5$, A.~Di~Virgilio$^1$, 
% A.~Gebauer$^{6,7}$, E. Maccioni$^{1,2}$,  A. Ortolan$^3$,
%  A.~Porzio$^{8,9}$, G.~Saccorotti $^{11}$, R.~Santagata$^{1,10}$,  A.~Simonelli$^{1,7}$,  and  G.~Terreni$^1$}
% \address{$^1$  INFN Sezione di Pisa, Pisa, Italy}           %% Put here the Institution's list
% \address{$^2$   Universit\`a di Pisa, Pisa, Italy}            %%
% \address{$^3$ LNL Laboratorio INFN, Legnaro, Italy}  
% \address{$^4$ DEI, University of Padova-Italy}
% \address{$^5$ INGV  Centro Nazionale Terremoti, L'Aquila, Italy}
% \address{$^6$ Technische Universitaet Muenchen,  Munich, Germany}
% \address{$^7$ Ludwig-Maximilians-University Munich, Germany}
% \address{$^8$  INFN Sezione di Napoli, Napoli, Italy}
% \address{$^9$  CNR-SPIN Napoli, Napoli, Italy}
% \address{$^{10}$   University of Siena, Siena, Italy}
% \address{$^{10}$   INGV sezione di Pisa, Pisa, Italy}

\author{Jacopo~Belfi}
\email{belfi@pi.infn.it}
\affiliation{INFN Sezione di Pisa, Pisa, 56127, Italy}

\author{Nicol\`o~Beverini}
\affiliation{Dipartimento di Fisica, Universit\`a di Pisa, Pisa, 56127, Italy}
\affiliation{INFN Sezione di Pisa, Pisa, 56127, Italy}

\author{Filippo~Bosi}
\affiliation{INFN Sezione di Pisa, Pisa, 56127, Italy}

\author{Giorgio~Carelli}
\affiliation{Dipartimento di Fisica, Universit\`a di Pisa, Pisa, 56127, Italy}
\affiliation{INFN Sezione di Pisa, Pisa, 56127, Italy}

\author{Davide~Cuccato}
\affiliation{Dipartimento di Ingegneria dell'Informazione, Universit\`a di Padova, Padova, 35131, Italy}

\author{Gaetano~De~Luca}
%\email{zhangjun@ustc.edu.cn}
\affiliation{INGV Centro Nazionale Terremoti, L'Aquila, 67100, Italy}

\author{Angela~Di Virgilio}
\affiliation{INFN Sezione di Pisa, Pisa, 56127, Italy}

\author{Andr\'e~Gebauer}
\affiliation{Department of Earth and Environmental Sciences, Ludwig-Maximilians University, Munich, 80333, Germany}

\author{Enrico~Maccioni}
\affiliation{Dipartimento di Fisica, Universit\`a di Pisa, Pisa, 56127, Italy}
\affiliation{INFN Sezione di Pisa, Pisa, 56127, Italy}

\author{Antonello~Ortolan}
\affiliation{Laboratori Nazionali di Legnaro-INFN, Legnaro, 35020, Italy}

\author{Alberto~Porzio}
\affiliation{CNR-SPIN Napoli, Napoli, 80126, Italy}
\affiliation{INFN Sezione di Napoli, Napoli, 80126, Italy}

\author{Gilberto~Saccorotti}
\affiliation{INGV Sezione di Pisa, Pisa, 56127, Italy}

\author{Andreino~Simonelli}
\affiliation{Department of Earth and Environmental Sciences, Ludwig-Maximilians University, Munich, 80333, Germany}
\affiliation{INFN Sezione di Pisa, Pisa, 56127, Italy}
\affiliation{Dipartimento di Scienze della Terra, Universit\`a di Pisa, Pisa, 56127, Italy}

\author{Giuseppe~Terreni}
\affiliation{INFN Sezione di Pisa, Pisa, 56127, Italy}

\date{\today}

\begin{abstract}
GINGERino is a large frame laser gyroscope investigating the ground motion in the most inner part of the underground international laboratory of the Gran Sasso, in central Italy. 
It consists of a square ring laser with a $3.6$ m side. Several days of continuous measurements have been collected, 
with the apparatus running unattended. The power spectral density in the seismic bandwidth is at the level of $10^{-10} \rm{(rad/s)/\sqrt{Hz}}$.
A maximum resolution of $30\,\rm{prad/s}$ is obtained with an integration time of few hundred seconds. The ring laser routinely detects seismic rotations 
 induced by both regional earthquakes and teleseisms. A broadband seismic station is installed on the same structure of the
gyroscope. First analysis of the correlation between the rotational and the translational signal are presented.
\end{abstract}

%insert suggested PACS numbers in braces on next line

\maketitle %\maketitle must follow title, authors, abstract and \pacs

%\section{Introduction}

\section{Introduction}    
Large Ring Laser Gyroscope (RLG) technology \cite{RSIUlli} provides very sensitive inertial rotation measurements. 
Among the most relevant recent results there are the direct observation of the rotational microseismic noise \cite{Hadziioannou2012} up to the detection of very long period 
geodetic effects on the Earth rotation vector \cite{ChandlerUlli}. The scientific community working on large frame RLGs had a rapid growth in the last decade. 
After the seminal work started in the '90s at the Canterbury University of Christchurch, (New Zealand) \cite{sitoNZ}, today, other laboratories around the world 
\cite{geosensor, dunn} both in Europe and US, use RLGs to detect ground 
rotational motions superimposed on the Earth rotation bias. A dedicated observatory of 3D seismic rotations, named ROMY \cite{sitoROMY}, 
was started this year in F\"urstenfeldbruck (Germany). The state of the art precision, is achieved by the Gross ring G \cite{G2009} in Wettzell (Germany), and it is better than some fractions 
of $\rm{prad/s}$, not far from $10^{-14} \rm{rad/s}$, that is the order of magnitude of the General Relativity effects in a ground based reference frame. 
The target of the GINGER (Gyroscopes IN GEneral Relativity) proposal is to measure the gravito-magnetic (Lense--Thirring) effect of the rotating Earth, by means of 
an array of high sensitivity RLGs \cite{NoiPRD}. Underground locations, far from external disturbances as hydrology, 
temperature and barometric pressure changes, are essential for this challenging experiments, and LNGS (Laboratori Nazionali del GranSasso, 
the underground INFN laboratory) may be a suitable one. %\textcolor{red}{as shown in Fig. \ref{topsoil} where the disturbances recorded by \textbf{G} in Wettzell are clearly visible \cite{NoiseUlli}}.
In order to test the local ground noise, a single axis apparatus called GINGERino, 
has been installed inside LNGS. This installation is a pilot-prototype for GINGER, and at the same time can provide unique information for geophysics \cite{simonelli}.
In addition, underground installations of large RLGs, free from surface disturbances, could provide useful informations to Geodesy \cite{EarthUlli}. 
Here the goal is to achieve a relative precision of at least 1 ppb in few hours of integration time, in order to integrate the information on Earth's rotation changes provided by 
the International Earth Rotation System (IERS) that, being based on the collection and elaboration of the observations of Very Large Base Interferometry (VLBI) and GPS systems, 
does not provide precise subdaily performance. The paper is organized as follows: in section \ref{apparatus} we describe the GINGERino optical and mechanical apparatus; section \ref{DAQ} is about the data acquisition
and transfer; section \ref{noise} contains the noise characterization and the illustration of the drift removal
method based on the backscattering noise identification and subtraction. In section \ref{seismology} we discuss some preliminary results on the seismic properties of the 
underground site as well as the analysis of the roto-translations induced by two far located earthquake events.  
Section \ref{conclusion} contains the conclusions of this work and the future perspectives of the experiment.
\section{GINGERino working principle and experimental setup}\label{apparatus}
RLGs measure rotation rate using the Sagnac effect. Oppositely propagating laser beams, generated inside a ring resonator
undergo a frequency splitting $\delta f$. For a horizontal RLG, located at colatitude $\theta$, the splitting $\delta f$ induced by the Earth's rotation rate $\Omega_E$, is expressed in function of  
the cavity area $A$, perimeter $P$ and laser-wavelength $\lambda$: 
\begin{equation}
 \delta f=\frac{4 A \Omega_E}{\lambda P}\cos(\theta+\delta\phi_{NS})\cos{\delta\phi_{EW}},
\end{equation} 
where $\delta\phi_{NS}$ and $\delta\phi_{EW}$ are the tilt angles, respectively in the North-South and in the East-West directions. 
GINGERino is a He-Ne laser operating on the red line at 633 nm. The square optical cavity, 3.6 m in sidelength, is made of four spherical mirrors with 4 m radius of curvature. 
The plane of the cavity of GINGERino is horizontal, thus the Sagnac frequency bias is provided by the projection of the Earth's rotation vector along the local vertical. 
At the latitude of LNGS the Sagnac frequency is $\rm{280.4\,\,Hz}$. 
%%%%%%%%%%%%%%%%
%%%%%%%%%%%%%%%%
The whole optical path of the beam inside the cavity is enclosed in a steel vacuum chamber, composed by 4 mirror chambers connected by vacuum pipes. 
The design is based on the GeoSensor design \cite{geosensor, NoiAPB, NoiVirgo}, where the alignment can be tuned by means of micrometric tip-tilt systems acting
on the mirror chambers orientation. From each corner of the cavity is possible to extract and detect the two counterrotating beams, 
so that the system has eight optical output beams. 
While monolithic cavities made of ultra-low expansion materials, like the Gross ring G in Wettzell, have an excellent passive stability,
they are not suited to form 3D arrays of very large size. On the other side, heterolithic systems require active stabilization, which is achieved by active control schemes
of the cavity geometry, exploiting very accurate optical wavelength references \cite{NoiCR, NoiRosa}. The present structure of GINGERino does not allow the 
full implementation of these techniques, and this sets a limit to its present long term stability.  
%GINGERino first installation uses the mechanical parts of the heterolithic G-Pisa ring laser \cite{NoiAPB, NoiVirgo}.
%It does not allow the full implementation of the full control of the cavity geometry, placing a limit to its present long term stability. 

\begin{figure}[!h]
 		\centering
 		\includegraphics[width=\linewidth]{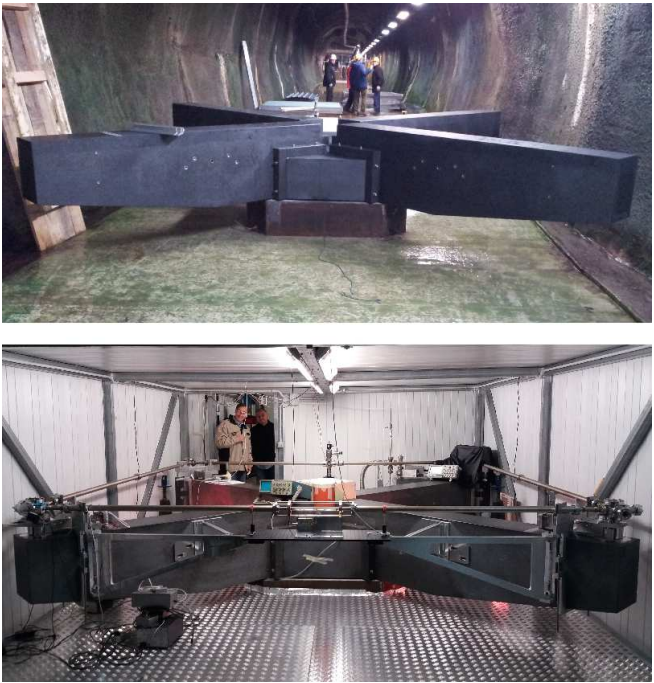} 
		\caption{Top: the granite frame of GINGERino, just after its installation in the LNGS tunnel. Bottom: completed setup inside the isolation chamber.}
		\label{construction}
\end{figure}

\begin{figure}[!h]
 		\centering
 		\includegraphics[width=\linewidth]{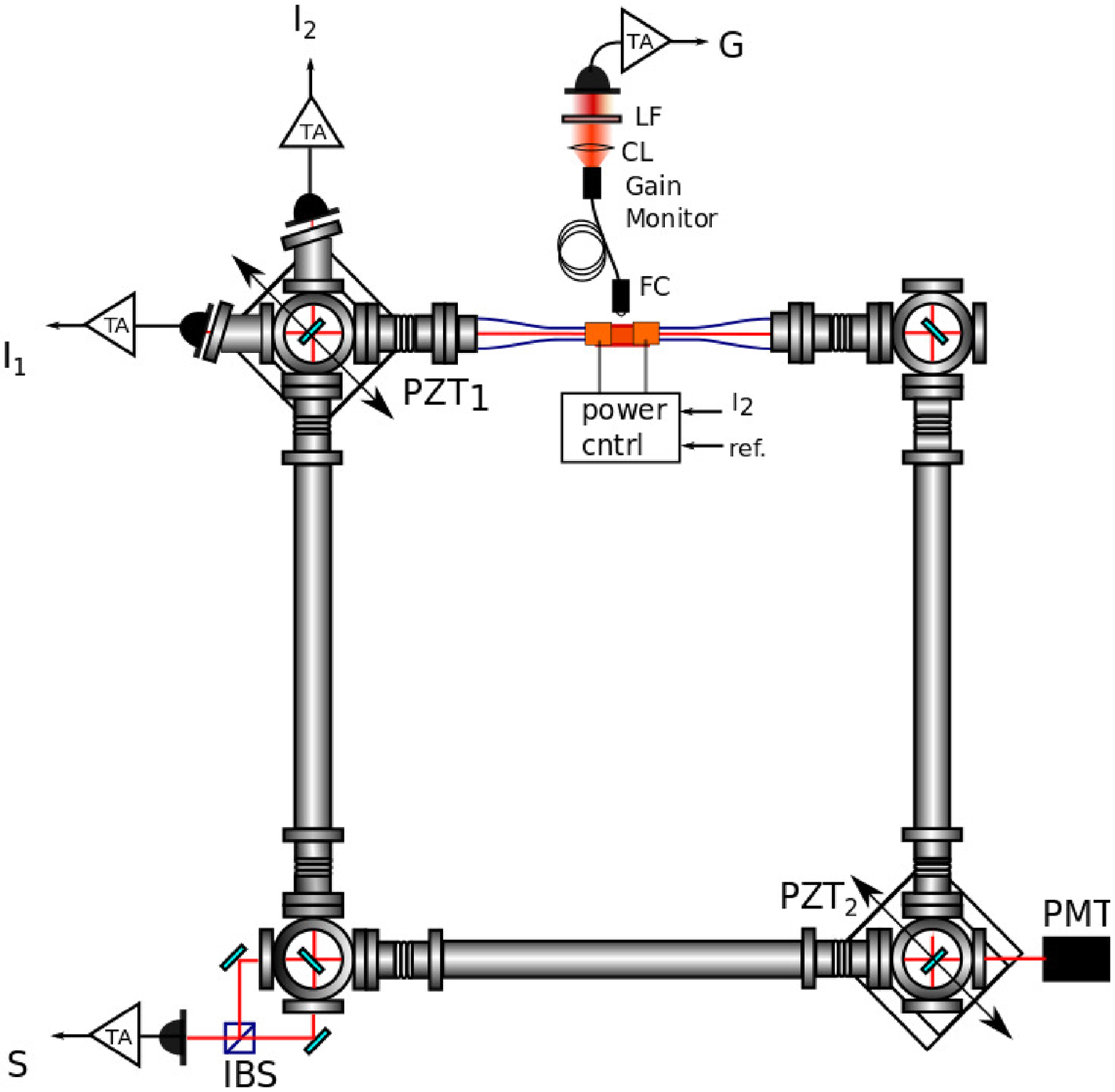} 
		\caption{Optical setup. Three optical signals are continuously acquired: the combined beams intensity (Sagnac interferogram) $S$ and the two monobeam intensities
		$I_{1}$ and $I_{2}$. The $G$ signal is the intensity of the plasma fluorescence, filtered around the laser line. It is acquired as a monitor 
		of the laser excitation level. IBS: Intensity Beam Splitter, PMT: Photo Multiplier Tube, LF: Line Filter (633 nm), TA:Transimpedance amplifier, FC=Fiber Coupler, CL: Collimating Lens.}
		\label{optical_apparatus}
\end{figure}

Mirrors dissipative losses set the shot noise limit to the sensitivity of a RLG, while the backscattering characteristics are responsible of the drift induced by the nonlinear 
coupling between the two counter-propagating laser beams (see ref.\cite{RSIUlli}). Dielectric deposition of thin films realized by very accurate ion beam sputtering procedures are typically applied and
 top quality substrates, with roughness of the order of fractions of angstrom, are necessary. State of the art dielectric mirrors can reach a reflectivity higher than $99.999\%$, with a total scattering of less than 4 ppm. 
These mirrors must be manipulated with the maximum care, possibly in clean environment (better than class 100) in order to avoid dust and humidity. A pyrex capillary with internal diameter of 4 mm is installed in the middle of one side and allows to excite the active medium (He-Ne plasma) 
by means of a radiofrequency capacitive discharge. The capillary diameter forces the laser to operate on single transversal mode ($\rm{TEM_{00}}$), 
while single longitudinal mode operation is obtained by keeping laser excitation near to threshold.  
Two piezoelectric translator stages can be used to stabilize the optical frequency of the laser against the cavity length variations induced by thermal expansion and
mechanical relaxations. This makes it possible to avoid laser mode hops, increasing the device duty cycle up to about 100$\%$. 
The four mirror chambers are tightly attached to a cross structure made of granite (african black), composed by a central octagonal massive block (3 tons), 
and four lightened arms each weighting $\approx \rm{800}\, \rm{kg}$ (see Fig.\ref{construction}). The granite structure is screwed 
to a reinforced concrete block anchored to the underneath bedrock. The African black granite has been chosen because it can be machined with high precision 
and has a low thermal expansion coefficient ($6.5\times10^{-6}$ $/^{\circ}C$). Being the whole set-up coupled to the ground in its central part, ground strain coupling to the cavity shape are minimized.
The installation area has a natural temperature of $8$ $^{\circ} C$ and a relative humidity close to the dew point all the year round. 
The whole installation is now protected by a large anechoic box. Infrared lamps are used to increase the temperature inside the 
box thus reducing the relative humidity from more of $90\%$ down to about $50-60\%$. We checked that no oscillations of temperature and humidity on 
the daily time scale are introduced by this method. Better isolation systems joined to a humidity control systems can be considered in the future.
So far, this infrastructure has been running for several months, and has shown that it keeps the GINGERino area at a 
temperature around $13$ $^{\circ}C$ with a stability better than $\rm{0.1}^{\circ}C$ for several days of operation. 
%%%%%%%%%%%%%%%%%%%%%%%%%%%%%%%%%%%%%%%%%%%%%%%%%%%%%%%%%%%%%%%%%%%%	
On top of the central part of the granite frame we installed additional instruments consisting in: one tiltmeter with nrad resolution (2-K High Resolution Tiltmeter (HRTM), Lipmann) 
and two high performance seismometers (Trillium 240s and Guralp CMG 3TÐ360s). 
The combination of different instruments is essential in the interpretation of the data and the characterization of the site. 
%%%%%%%%%%%%%%%%%%%%%%%%%%%%%%%%%%%%%%%%%%%%%%%%%%%%%%%%%%%%%%%%%%%%
In Fig.\ref{optical_apparatus} is sketched the optical scheme of GINGERino in the present configuration. Three transimpedance amplified silicon photodiodes, with a bandwidth of 
4 kHz, are used to detect the Sagnac interference signal at 280.4 Hz and the two single beam intensities. A photomultiplier (PMT), with a bandwidth of 400 MHz, is used for a double purpose: 
to detect the radiofrequency signals produced by the beating between higher order lasing modes and to occasionally perform ring-down time measurements for estimating the intracavity losses.
\section{Data acquisition}\label{DAQ}
GINGERino runs unattended, in this way man made disturbances are minimized. We have developed a remote 
interface with the experiment that allows us to monitor the status of the apparatus and also to drive the mirror positioning PZT actuators sketched in Fig. \ref{optical_apparatus}.
The DAQ system itself is remote-controlled and transfers the data from INFN-LNGS to INFN-Pisa. 
The DAQ hardware has been selected in order to be transportable; it is based on the PXI-8106 controller by National Instruments.  Its main tasks can be listed as follows (referring to Fig. \ref{optical_apparatus}):
\begin{itemize}
\item analog to digital conversion and storage of the Sagnac signal $S$ and the two mono beams signals $I_{1,2}$ with $5$ kHz sampling rate;
\item analog to digital conversion and storage of environmental signals (temperature, humidity, pressure), laser parameters 
(plasma fluorescence gain monitor $G$, average intensities, piezoelectric transducers driving voltage) and tiltmeters, with $\rm{1\,\, Hz}$ sampling rate;
\item real-time processing of experimental parameters connected to laser gain, backscattering parameters, actuators signals required by active control loops;
\item digital to analog generation of the signals driving the laser, necessary for some of the controls of the apparatus.
\end{itemize}
Acquired data are written in the PXI local hard-disk. Both frequency and time accuracy are important since the former affects the estimation of the Sagnac frequency and the latter introduces errors in the synchronization 
of the RLG data stream with the data streams of other instruments (mainly seismometers). The PXI receives a GPS-synchronized PPS (pulse per second) signal and is connected to a local NTP server in order to obtain a time stamp with the required precision.
The frequency accuracy is obtained by disciplining the 10 MHz clock of the PXI-6653 board to the PPS via the PXI-6682. The error on the time stamp is limited by the uncertainty on the NTP, which is of the order of a few milliseconds. The data written on the PXI hard-disk are copied via FTP into 
a dedicated directory on a local virtual machine and then copied into the final data storage destination (at INFN Pisa). 
\section{Sensitivity of the apparatus}\label{noise}
\begin{figure}[!h]
 \centering
 \includegraphics[height=5.8 cm]{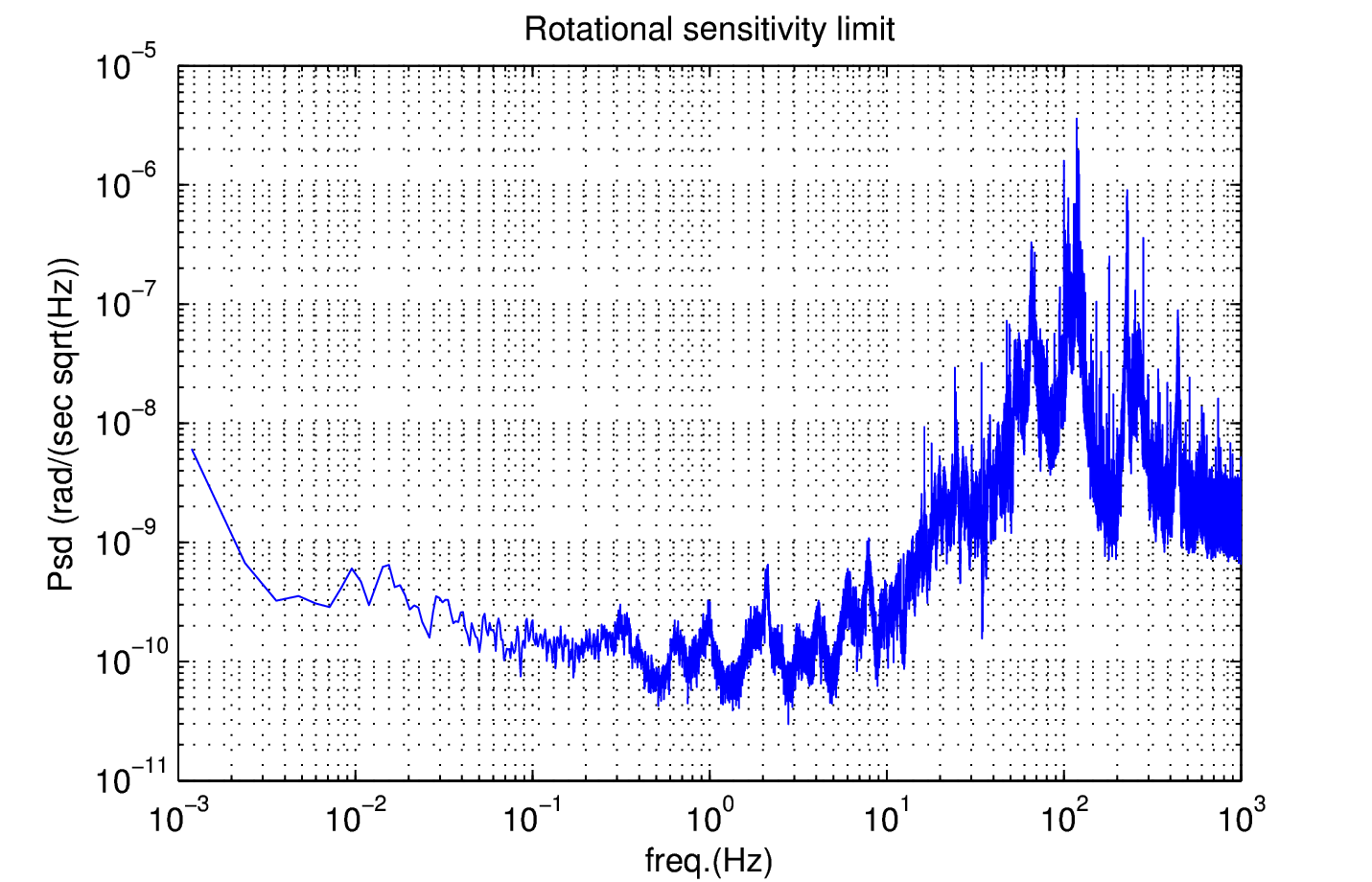} 
 \caption{Angular velocity linear spectral density of GINGERino calculated a dataset of 3600 s from 11-06-2016 at 00:00. Power spectral density is estimated from the raw data interferogram.}
 \label{PSD0}
 \end{figure}

From a direct estimate of the Sagnac frequency by means of the Hilbert transform of the interferogram, 
we deduced an instrumental sensitivity limit at the level $\rm{100\, prad/sec/\sqrt{Hz}} $ in the range $(10^{-2}- 1) \rm{Hz}$. 
A typical rotational noise spectrum is shown in Fig.\ref{PSD0}. Data refer to 1 hour acquisition on the 11th of June 2016. 
For this run the cavity ringdown time was $\sim 250\,\mu s$ corresponding to a total loss per round trip of about $\rm{190\,ppm}$. 
As clearly visible, the long term stability of the raw data is limited to 10-20 s, mainly by radiation backscattering on the mirrors.
 A reduction of the backscattering induced frequency noise can be obtained by identifying and subtracting its 
 contribution from the measurements of the single beam outputs from the cavity. This has been done for the run of June 2016, and it is discussed in the next section.
\section{Backscattering analysis}
\begin{figure}[!h] 
\centering
\includegraphics[width=7 cm]{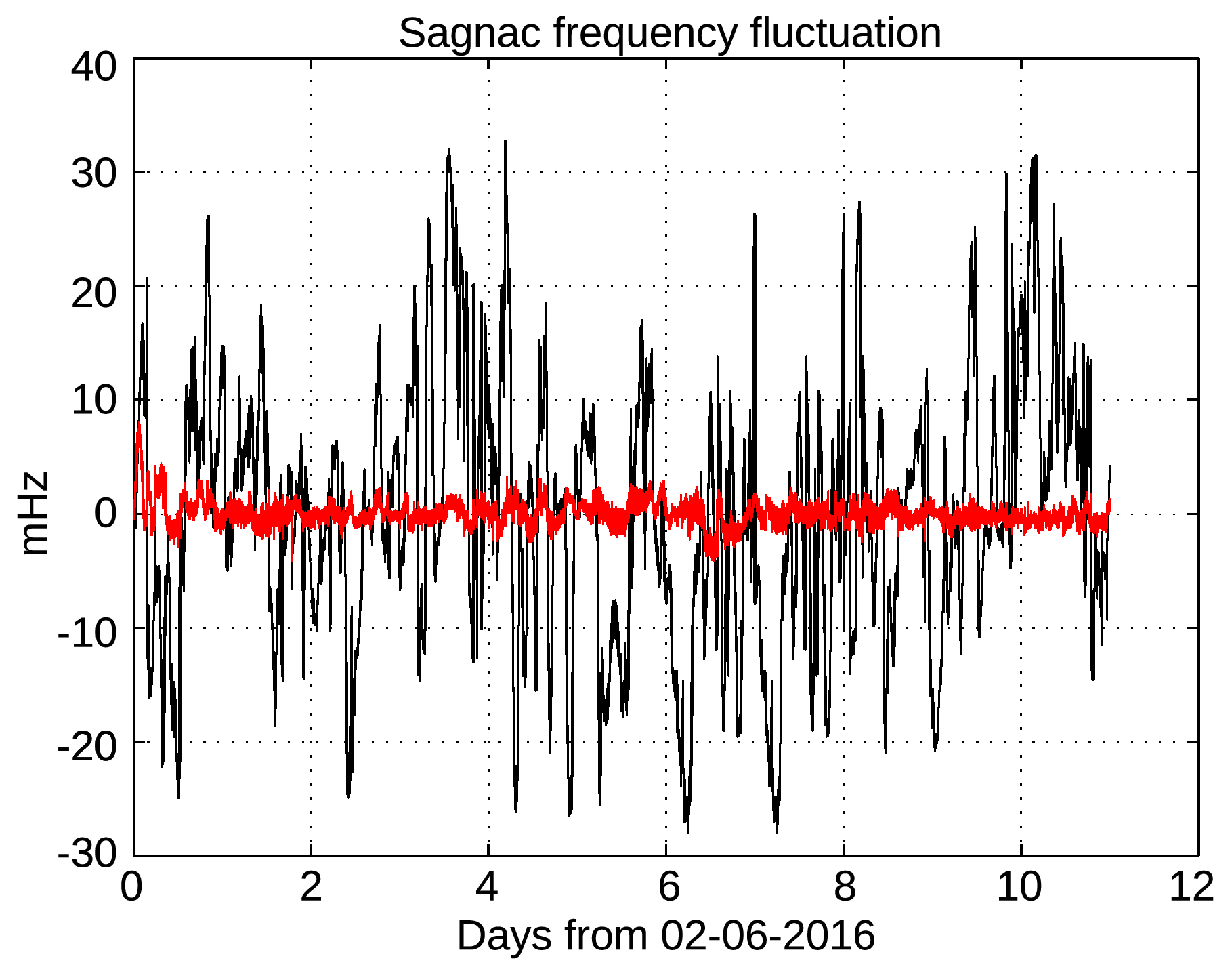}
\caption{Black: raw data. Red: backscattering corrected data.}
\label{risu}
\end{figure}

The strategy for subtracting backscattering noise from ring-laser data  has been extensively discussed in \cite{NoiAO} and  \cite{NoiMetrologia}, 
where we have shown how and why backscattering noise can be efficiently subtracted, by applying an Extended Kalman Filter (see \cite{NoiMetrologia} for details).
The time dependence of backscattering contribution can be also estimated using  a  model which assumes reciprocal ring laser parameters. 
This approach has been exploited in \cite{Hurst:14}, where the backscattering parameters were estimated by fitting amplitudes and phases of the two monobeam intensities. 
It has been tested that the two methods give similar results.  
For this analysis purpose several service signals are necessary: the two monobeam intensities and the laser gain. 
Data were processed following the procedure already developed for G-Pisa \cite{NoiVirgo}, tuning the pre-filters  to the GINGERino Sagnac frequency, 
and estimating the laser parameters by averaging over 10 seconds the mono-beams intensities. We firstly extract from the data mono-beam intensities, 
modulations and phase differences, then we use these quantities to estimate the laser parameters connected to backscattering at a rate of 1 sample every 10 seconds. 
After the parameter identification, backscattering contribution is 
calculated. Results for a time series of 12 days are shown in Fig.\ref{risu}. 
 %%
\begin{figure}[!h]  \centering
\includegraphics[width=8 cm]{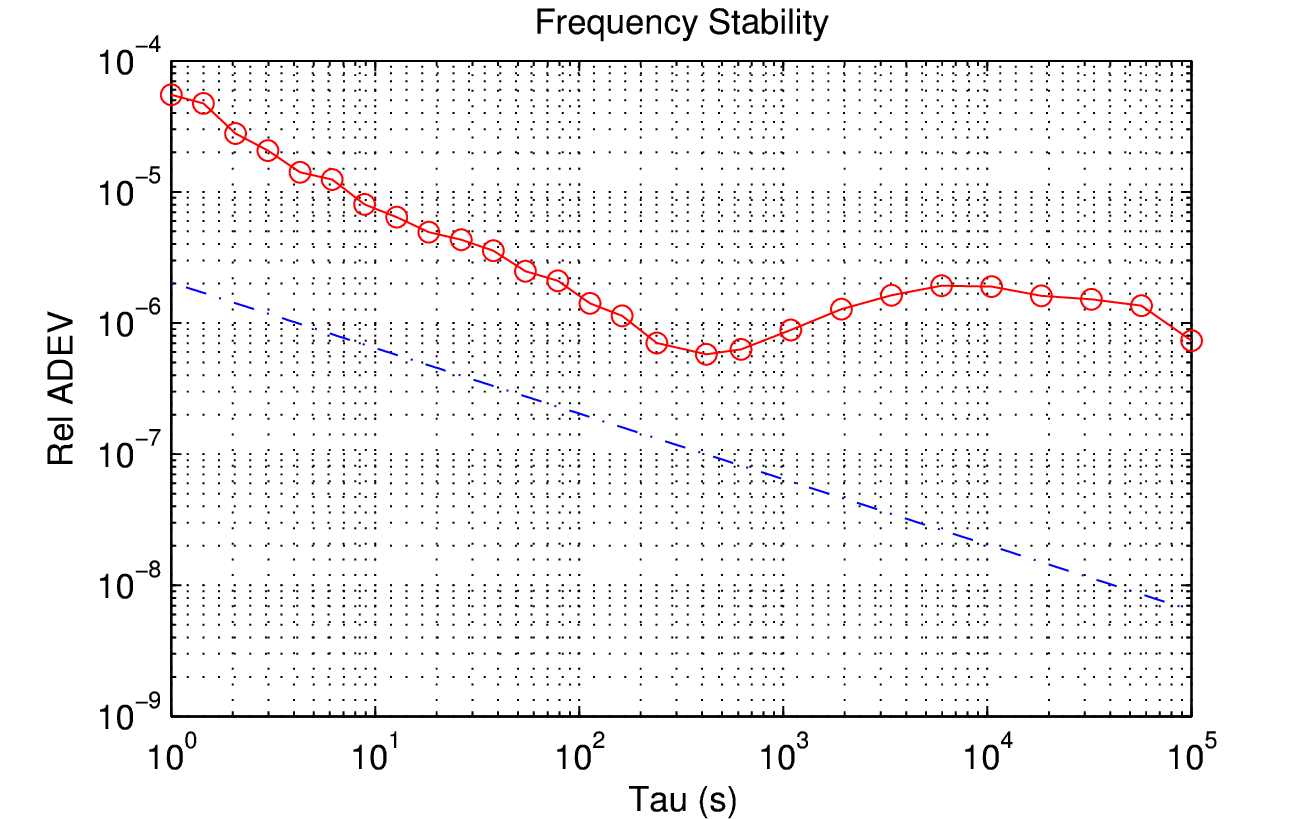} 
\caption{Relative Allan deviation for the Sagnac frequency after backscattering subtraction. Straight line represents the calculated shot noise limit.}
\label{corr_allan}
\end{figure} 
The relative Allan deviation of the backscattering corrected data is shown in Fig.\ref{corr_allan}.
\section{Seismological observations}\label{seismology}
\begin{figure}[!h]
    \includegraphics[height=5.6 cm]{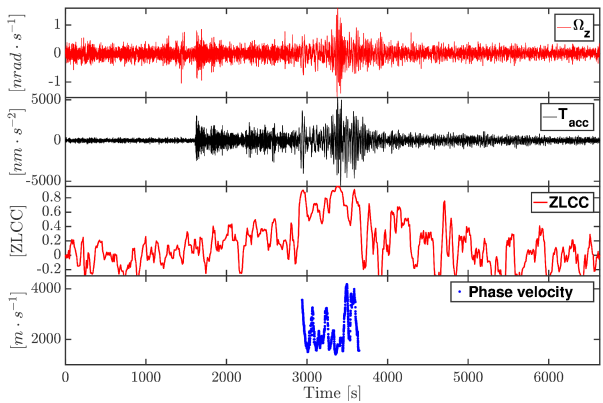} 
     \caption{Mid Atlantic ridge earthquake, June 17, 2015, 12:51 p.m., MWC 7. Top: seismograms for the transverse acceleration (black) and vertical rotation (red). 
    Center: the zero lag correlation coefficient between rotation and transverse acceleration. Bottom: apparent phase velocity of the surface waves. 
    Phase velocities are computed for the seismograms parts where the correlation between rotation and translation is larger than 0.6.}
  \label{mid}
  \end{figure}

\begin{figure}[!h]
  \includegraphics[height=5.6 cm]{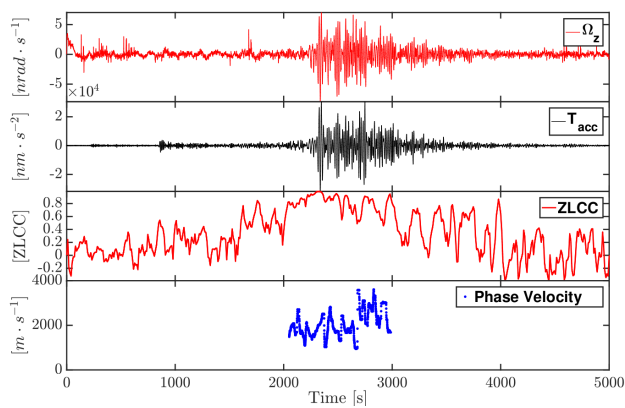} 
    \caption{Rykyu Islands earthquake,  November 13, 2015, 08:51 p.m., MWC 6.8. Plot legend is the same as in Fig.\ref{mid}.}
    \label{jap}
  \end{figure}

The two independent digitizers for the RLG and the seismometers are synced to the GPS time reference. This allows the direct comparison between 
rotational and translational signals \cite{Igel01}. Two teleseismic events are reported in the following. Results are shown in Fig.\ref{mid} and \ref{jap}. The upper two traces indicate the time history of 
transverse acceleration and  rotation rate as detected by the seismometer and RLG, respectively. The N and E components of the seismometer, after being corrected for the instrumental response, 
have been rotated in order to construct the transverse component which is analyzed in comparison with the gyroscope signal. We evaluated the Zero-Lag Correlation  
Coefficient (ZLCC) between transverse acceleration and rotation using a time window of 50 seconds, sliding with 50$\%$ overlap. 

%%%%%%%%%%%%%%%%%%%%%%%

For Love seismic waves \cite{id415}, in the plane wave approximation, the transverse acceleration and the vertical rotation signals are in phase and their ratio is  
proportional to the phase velocity. Phase velocity measurements contain information about the elastic properties of the ground, 
and are typically obtained by means of seismometer arrays installations. A system composed by a horizontal RLG and a seismometer 
provide the same information in a single site installation. An example of this estimate is given in lowest plot of Fig.\ref{mid} and \ref{jap}.

%%%%%%%%%%%%%%%%%%%%%%%

\begin{figure}[!h]
  \includegraphics[width=\linewidth]{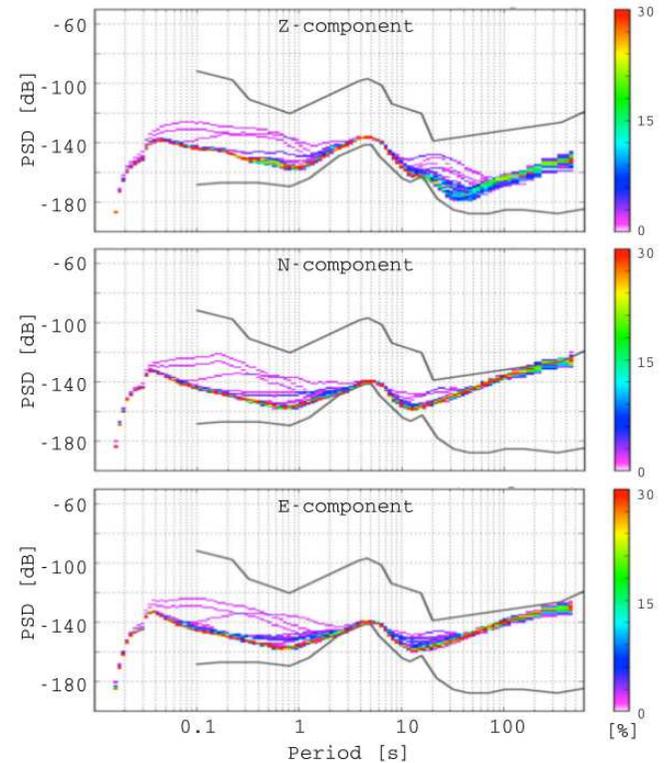} 
    \caption{Probabilistic power spectral densities for the three components of ground acceleration as recorded by the seismometer. Vertical scale is relative to $\rm{1\,m^2 s^{-4} Hz^{-1}}$. 
    For each frequency bin, the maps illustrate the probability of  observing a given spectral power, according to the color scale at the right.}
    \label{seismometers}
  \end{figure}

Surface wave phase velocity is calculated in the two cases for the points where the ZLCC is above 0.6. 
The characteristics of the background seismic  noise at the site are illustrated in Figure  \ref{seismometers}, where the probabilistic power spectral densities (Pround PSD) of the three 
components of ground acceleration are compared to the High- and Low-Noise Models (NHNM and NLNM, respectively) \cite{Peterson}. 
The typical spectra are close to the NLNM and shows a very good behavior throughout the spectral region of the primary and secondary micro-seism (i.e., at periods spanning the 1-10s interval), 
exhibiting however larger and unwanted noise at low frequency (long periods) for the N and E components.
\begin{figure}[!h] 
    \centering
    \includegraphics[width=\linewidth]{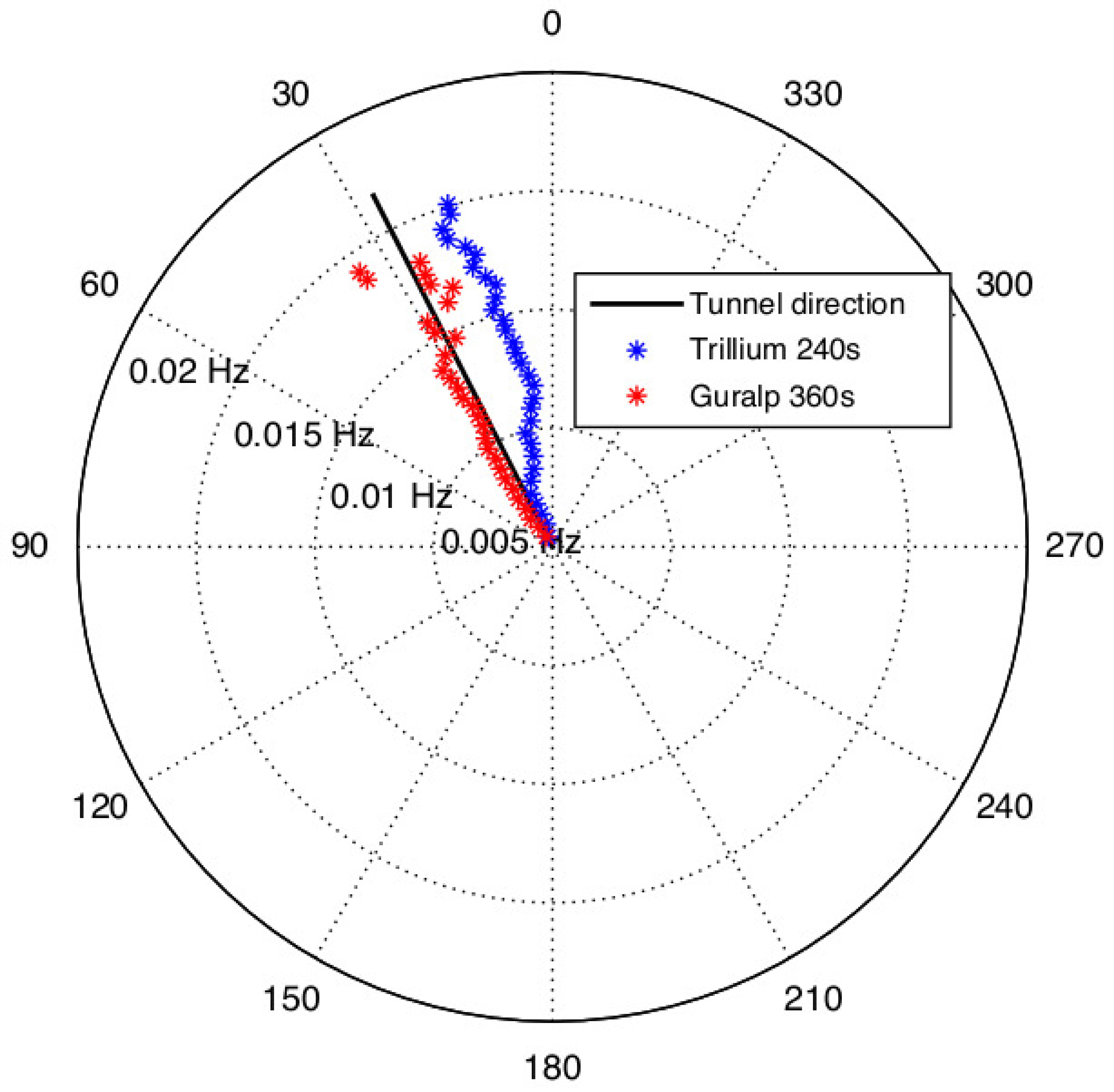} 
% \caption{Reconstruction of the polarization of the horizontal noise of the two seismometers. Their relative alignment is within 5 degrees, 
% they show the disturbances propagates mainly along the tunnel.}
  \caption{Principal polarisation direction of the background noise over the $\rm{50-200\,s}$ period range. 
  Results from the two seismometers are coherent within a 5 degrees tolerance, indicating ground oscillations aligned along the tunnel's direction. 
  This suggests a main control of the underground cavity in the generation of seismic noise at very long periods.}
  \label{polar}
\end{figure}
A deeper analysis consisting in the calculation of the noise polarization over the horizontal plane shows that the
noise polarization is markedly directional and directed along the tunnel (see Fig.\ref{polar}). Accordingly to the literature \cite{Beauduin}, 
a possible explanation is that the long-period, high-amplitude noise is induced by the conveyed air motion in the tunnel.  
 \section{Conclusions}\label{conclusion}
GINGERino has been constructed inside LNGS and performs ground rotation measurements with a very high duty cycle. The system provides Earth rotation rate measurements as well as seismic rotational 
data thanks to a dedicated architecture for laser remote control, data acquisition and data transfer. The sensitivity curve shows a level around $\rm{10^{-10}\,rad/s}$ 
compatible with the actual instrument shot noise and ringdown time.
During the first runs all the major teleseismic events present in the Global CMT Catalog have been detected. The standard rotation/transverse-acceleration correlation analysis 
is presented for two different events. Long term stability of raw data is limited by backscattering noise, which can be subtracted in large part via post processing. After correcting the backscattering induced drift
a maximum resolution of about $30\, \rm{prad/s}$ for 500 s of integration time is obtained. The correlation between the observed instabilities of the gyroscope and the 
environmental parameters fluctuation (temperature, pressure, humidity, anthropic activities) is under investigation.  
\section*{Acknowledgement}
The construction of GINGERino has been possible thanks to the effort of several colleagues of Pisa, Naples and LNGS. For that we are grateful to:
A. Soldani, G. Petragnani, G. Balestri, M. Garzella, A. Sardelli and F. Francesconi from Pisa, and   G. Bucciarelli, L. Marrelli 
and N. Massimiani (Servizio Esercizio e Manutenzione) and F. Esposito, L. Faccia, G. D'Alfonso, S. Giusti and 
L. Rossi of 'Servizio Facchinaggio' at LNGS. We thank U. Schreiber for his constant help on ring lasers issues and H. Igel and J. Wassermann for their support on geophysical aspects of the research. 
Special thanks go to C. Zarra for her continuous assistance at LNGS and to Pino Passeggio of INFN section of Naples, 
who has taken care of the installation of the GINGERino chamber. We have to thank as well the Computing and Network teams of 
LNGS and Pisa Section for their continuous assistance. We have to acknowledge Laurent Pinard of 
the Laboratoire des Mat\'eriaux Avanc\'es (CNRS-Lyon), who has  given us the necessary support for everything concerning the mirrors.
\section*{References}
\bibliography{mybibRSI}
\bibliographystyle{unsrt}
\end{document}